\documentclass[12pt]{article}

\usepackage{cite}
\usepackage{epsf}
\usepackage{graphicx}
\usepackage{psfrag}
\usepackage[dvips]{epsfig}
\usepackage[dvips]{graphicx}

\usepackage[english]{babel}

\usepackage{amssymb,indentfirst}
\usepackage{amsmath}

\usepackage{verbatim}

\makeatletter
\renewcommand \thesection {\@arabic\c@section.}
\renewcommand\thesubsection   {\thesection\@arabic\c@subsection.}
\renewcommand\thesubsubsection{\thesubsection\@arabic\c@subsubsection.}
\makeatother

\begin{document}
\title{Lower Bound on the Magnetic Field Strength of \\ a Magnetar from Analysis of SGR Giant Flares
}

\author{
 A.~A.~Gvozdev\footnote{E-mail: gvozdev@uniyar.ac.ru}, \,
I.~S.~Ognev  \footnote{E-mail: ognev@uniyar.ac.ru}, \,
E.~V.~Osokina \\
{\small Yaroslavl State University, Sovietskaya 14,}\\
{\small 150000 Yaroslavl, Russia.}}
\date{}
\maketitle


\begin{abstract}
\noindent
Based on the magnetar model, we have studied in detail the processes of neutrino cooling of an
electron--positron plasma generating an SGR giant flare and the influence of the magnetar magnetic field
on these processes. Electron--positron pair annihilation and synchrotron neutrino emission are shown to
make a dominant contribution to the neutrino emissivity of such a plasma. We have calculated the neutrino
energy losses from a plasma-filled region at the long tail stage of the SGR 0526--66, SGR 1806--20, and
SGR 1900+14 giant flares. This plasma can emit the energy observed in an SGR giant flare only in the
presence of a strong magnetic field suppressing its neutrino energy losses. We have obtained a lower bound
on the magnetic field strength and showed this value to be higher than the upper limit following from an
estimate of the magnetic dipole losses for the magnetars being analyzed in a wide range of magnetar model
parameters. Thus, it is problematic to explain the observed energy release at the long tail stage of an SGR
giant flare in terms of the magnetar model.

\vspace{5mm}


\end{abstract}




\subsection*{INTRODUCTION}

Soft gamma-ray repeaters (SGRs) and anomalous
X-ray pulsars (AXPs) constitute a special class
of neutron stars with anomalously large spin periods $P~\sim~(5~-~8)s $ and spindown rates $\dot P \sim (10^{-10} - 10^{-13})s\, s^{-1}$. About 20 such objects have been discovered
to date in our and nearest galaxies (Mereghetti
2008). Their identification with supernova remnants
shows them to be young isolated neutron stars
with ages $\tau_{NS} \sim (10^3 - 10^4)$ yr without accretion disks
(Bisnovatyi-Kogan 2006). Note the most characteristic
properties of these objects. First, in quiescence
they have an anomalously high effective temperature
for isolated stars, emitting soft X rays from the
surface with luminosities $L_{NS} \sim (10^{33} - 10^{36})$ (Mereghetti 2008). Second, AXPs exhibit large
glitches (Dib et al. 2008), whileSGRs exhibit gamma-ray
bursts (Strohmayer and Watts 2006; Watts
and Strohmayer 2007). They are interpreted as a
manifestation of a seismic activity in these objects
similar to the seismicity of the Earth, the Sun, and
the young Vela pulsar (Gogus et al. 2000). Third,
gamma-ray bursts were also detected from AXPs,
although they are not so powerful as those from SGRs
(Woods et al. 2005). This suggests that SGRs
and AXPs most likely belong to the same class of
neutron stars. The SGR flare activity manifests itself
in the emission of numerous (up to 100 episodes
per day) short bursts in the energy range from hard
X rays to soft gamma rays with a typical duration
of
$ \tau_F \sim 0.1$  and energy $E_F \lesssim 10^{41} $ erg. In several
cases, a series of short bursts was followed by a giant
flare exceeding in energetics the short ones by several
orders of magnitude. Two easily distinguishable
stages were observed in the three most powerful
(in energy release) giant flares from SGR 0526--66
(March 5, 1979), SGR 1900+14 (August 27, 1998),
and SGR 1806--20 (December 27, 2004): a short,
with a duration $\tau_{HS} \sim (0.25 - 0.5)$ and energy $E_{HS}  \sim  (10^{44}  - 10^{46})$ erg,  hard spike (HS) followed by a pulsating
long $(\tau_{LT} \sim (200 - 400) )s $ tail (LT) with energy
$E_{LT} \sim (1 - 4) \times 10^{44}$  erg,
at which a modulation of
the emission intensity by the neutron star spin period
was observed. Here, we investigate the energy losses
at the LT stage of an SGR giant flare and do not
consider the HS stage. Below, we present data for the
three most energetic SGR giant flares at this stage:
\\
SGR~0526-66 ($ D \approx 55$~kpc) $ \tau_{LT} \approx 200 $~s, $ E_{LT} \approx 3.6\times 10^{44} $~erg
(Mazets et al. 1979); \\
SGR~1900+14 ($D \approx 15$~kpc) $ \tau_{LT} \approx 400 $~s, $ E_{LT} \approx 1.2\times 10^{44} $~erg
(Ibrahim et al. 2001); \\
SGR~1806-20 ($D \approx 15$~kpc) $ \tau_{LT} \approx 380 $~s, $ E_{LT} \approx 1.3\times 10^{44}$~erg (Mereghetti et al. 2005; Frederiks et al. 2007). As can be seen, these parameters
almost coincide, suggesting a unified flare formation mechanism at the LT stage.

Note that the energy release in SGR giant flares
is smaller than that in supernova explosions and
cosmological gamma-ray bursts only. The model
of a magnetar, a neutron star with an anomalously
strong magnetic field $B_M \sim  10^{15}$ G (Duncan and
Thompson 1992; Thompson and Duncan 1993),
was proposed to explain such a huge gamma-ray
burst energy for an isolated neutron star. In this
model, the energy of the magnetic field liberated when
its configuration changed rapidly is assumed to be
the gamma-ray energy source. Subsequently, the
magnetar model was used to describe the emission
from AXPs and SGRs in quiescence (Thompson
and Duncan 1996) and during an SGR giant flare
(Thompson and Duncan 1995). In the magnetar
model of an SGR giant flare (Thompson and Duncan
1995), the long-term  $( \tau_{NS} \sim 10^3 yr )$ evolution
of the magnetar core with poloidal and toroidal
magnetic fields of strength $ B_M \sim 10^{15}$~G is assumed
to be ended with a starquake leading to a largescale
plastic deformation of its crust. The electric
currents emerging during the deformation produce a
perturbative magnetic field with field lines closed on
the crust. The region with closed field lines is rapidly
(in hundredths of a second) filled with an electron--
positron plasma trapped by this field (the so-called
fireball). A fairly hot plasma, with a temperature $T \sim 10^{10}$~K, is generated.
The X-ray photon flux observed
at the LT stage is assumed to be emitted from a
thin near-surface layer of the fireball. Thompson and
Duncan (2001) modeled this emission and compared
it with the observed light curve of the SGR 1900+14
giant flare. As a result, the distributions of plasma
parameters (temperature and magnetic field) that
agreed best with the observational data were obtained.
It is important to note that the neutrino
emission was disregarded by these authors, because
it was assumed to be significantly suppressed by a
strong magnetic field inside the fireball.

Here, we study in detail the plasma neutrino emission
processes at the LT stage based on the magnetar
model of an SGR giant flare. We show that
the plasma energy losses through neutrino emission
are significant even in the case of a strong magnetic
field with a strength  $ B \gtrsim 10^{15}$  G. For the most energetic
flares from SGR 0526--66, SGR 1806--20,
and SGR 1900+14, we model the fireball neutrino
cooling. The dependences of neutrino cooling on parameters
of the temperature and magnetic field distributions,
fireball size, and its total energy are analyzed.

Below, except for the specially
stipulated cases, we use a system of units in which $c = \hbar = k = 1 $.


\subsection*{ \bf
NEUTRINO COOLING OF A RELATIVISTIC \\
NONDEGENERATE ELECTRON--POSITRON PLASMA }

\subsection*{\it
Main Neutrino Processes}

The following reactions are the most significant
neutrino emission processes of a relativistic nondegenerate
electron--positron plasma: the electron--positron pair annihilation into a pair of neutrinos with
an arbitrary flavor
\begin{equation}
 e^-  + \, e^+ \to \nu_i + \tilde\nu_i  ,
 \label{e^+e^-}
\end{equation}
the plasmon decay into a neutrino pair
\begin{equation}
\gamma \to \nu_i + \tilde\nu_i ,
\label{gamma pl}
\end{equation}
the neutrino production due to the fusion of two photons
\begin{equation}
\gamma + \gamma \to \nu_i + \tilde\nu_i ,
\label{gamma+gamma}
\end{equation}
the photoneutrino emission process
\begin{equation}
e^\mp + \gamma \to e^\mp + \nu_i + \tilde\nu_i ,
\label{emp1}
\end{equation}
the neutrino synchrotron emission by electrons
\begin{equation}
e^\mp \overset{B}{\to} e^\mp + \nu_i + \tilde\nu_i ,
 \label{e^mp}
\end{equation}
which is kinematically allowed only in an external
magnetic field. Here and below, the subscript  $ i = e,\mu, \tau $ specifies the neutrino flavor.

\subsection*{ \it
The Case of a Weak Magnetic Field}

\noindent

We will begin our analysis of the neutrino emission
processes with the case of an electron--positron
plasma in the absence of a magnetic field. The neutrino
emissivity of an ultrarelativistic nondegenerate
plasma in the annihilation process (1) is well known
(Kaminker et al. 1992) and can be written as
\begin{equation}
 Q^{(0)}_A
= \frac{7 \, \zeta(5) \, C_{+}^{2}} {12 \, \pi} \, G^2_F \, T^9.
\label{QA0}
\end{equation}
Here, $T$~--is the plasma temperature,  $G_F$~--is the Fermi
constant, $ C_{+}^{2} = \sum_i \left ( c^2_{v_i} + c^2_{a_i} \right ) \simeq 1.675 $,
where $c_{v_i}$ and $c_{a_i}$ are the vector and axial constants of the leptonic electroweak current, and  $\zeta (x)$  is the Riemann zeta function.

The emissivity in the plasmon decay into a neutrino
pair~(\ref{gamma pl})  has also been well studied and can
be found, for example, in Yakovlev et al. (2001) and
Kantor and Gusakov (2007). For a nondegenerate
plasma, it can be represented as
\begin{equation}
Q^{(0)}_{\rm P} =
\frac{C_{v}^{2}} {324 \, \pi}
\, \alpha^2 \, G^2_F \, T^9,
\label{Q pl}
\end{equation}
where $\alpha = 1/137$~--is the fine-structure constant and $C^{2}_{v} = \sum_i c_{v_i}^2 = 0.9248$.
It is easy to see that the emissivity
in this process is strongly suppressed compared
to the electron--positron pair annihilation.

The processes~(\ref{gamma+gamma}) and~(\ref{emp1}) in a plasma were extensively
studied previously (Beaudet et al. 1967; Itoh
et al. 1996) and the neutrino emissivities in these processes
were shown to be also negligibly small compared
to (6). Thus, the annihilation process makes
a major contribution to the neutrino emissivity of a
relativistic nondegenerate plasma.

In the case of a relatively weak magnetic field, $T^2 \gg eB$ the emissivities in the processes  (\ref{e^+e^-}), (\ref{gamma pl}), (\ref{gamma+gamma})  and~(\ref{emp1}) change insignificantly, but the presence
of a magnetic field makes the new synchrotron neutrino
pair production process ~(\ref{e^mp}) kinematically open.
The emissivity in this process in the limit of a weak
magnetic field is given by the expression (Kaminker
and Yakovlev 1993):
\begin{equation}
Q^{(0)}_S = \frac{10 \, \zeta(5)} {9 \, (2 \pi)^5}
\, C_{+}^{2} \, G^2_F (eB)^2 \, T^5
\left [ \ln \left (\frac{T^2}{eB} \right ) + 4.66 \right ],
\label{QS}
\end{equation}
from which it follows that $Q^{(0)}_S / Q_A^{(0)} \sim (eB / T^2)^2$ i.e., the emissivity in this process in the limit under
consideration is also suppressed. Thus, in a relatively
weak magnetic field, the $e^\pm$ pair annihilation reaction
is the main plasma neutrino cooling process and the
total neutrino emissivity is defined by Eq.~(\ref{QA0}).

The above analysis allows the characteristic plasma
neutrino cooling time~$\tau^{(0)}_\nu$
to a temperature~$T$ to be calculated.
It can be found from the equation
\begin{equation}
\frac{d}{dt}\left[ \frac{11 \pi^2}{60} T^4 \right] = - Q_A^{(0)},
\label{eq:cool}
\end{equation}
where the total energy density of an ultrarelativistic
nondegenerate $e^\pm $ plasma and photons appears on
the left-hand side. The solution of this equation,
which essentially coincides with the estimate from
Thompson and Duncan (1995)
\begin{equation}
\tau^{(0)}_\nu \simeq \frac{44 \, \pi^3}{175 \, \zeta (5) \, C_{+}^{2}} \,
\frac{1}{G_F^2 \, T^5} \simeq 22~\mbox{s} \,
\left ( \frac{1~\mbox{MeV}}{T} \right )^5 ,
\label{eq:tau-nu}
\end{equation}
shows that the neutrino cooling time for a fairly
weak magnetic field is an order of magnitude shorter
than the characteristic duration of a giant flare
$\tau_{LT} \simeq (200 - 400)$~s. Thus, the bulk of the hot-plasma energy
is expended in cooling by neutrino emission and such
a plasma cannot be the source of an SGR giant flare.
Consequently, a mechanism suppressing the neutrino
cooling processes is required. As will be shown below,
a strong magnetic field that is capable of significantly
reducing the hot-plasma energy losses through neutrino
emission can act as such a mechanism.

\subsection*{
\it
The Case of a Strong Magnetic Field}

In the asymptotic limit of a strong magnetic field,
when $eB \gg T^2 \gtrsim m^2$ , the emissivity in the electron--positron pair annihilation process (1) is well known
and can be represented for a nondegenerate plasma
as (Kaminker et al. 1992)
\begin{equation}
Q^{(B)}_A = \frac{\zeta (3) \, C_{+}^{2}} {48 \pi^3}
\, G^2_F \, m^2 \, eB \, T^5 ,
\label{QAB}
\end{equation}
where $m$ is the electron mass.

The plasmon decay reaction~(\ref{gamma pl}) is also modified
significantly by a strong magnetic field, because not
only the amplitude of the process but also the dispersion
law of plasmon modes change. Calculations
show (Kuznetsov et al. 1998) that the emissivity in
the strong-field limit is
\begin{equation}
Q_{P}^{(B)} = \frac{\zeta (5) \, C_{+}^{2}}{2 \pi^6}
\, \alpha \, G^2_F \, (eB)^2 \, T^5 .
\label{Q gamma}
\end{equation}
As we see from this expression, the emissivity in this
process can become equal to (11) only at a sufficiently
large magnetic field strength $B \simeq 8 \times 10^{15}$~G.

The fusion of two photons  (\ref{gamma+gamma}) in the case of a
strong magnetic field was investigated by Rumyantsev
and Chistyakov (2008). Since the analytical
expression for the neutrino emissivity in this reaction
is fairly cumbersome, it is given below in an
approximate form that is valid only in the limit $eB \gg T^2 \gg m^2$:
\begin{equation}
Q_{\gamma\gamma}^{(B)} \simeq
2.7 \times 10^{18}
\frac{\mbox{erg}}{\mbox{cm}^3 \, \mbox{s}}
\left ( \frac{T}{m} \right )^9 .
\label{Q gamma 110}
\end{equation}
Note that the emissivity in this limit is virtually independent
of the magnetic field strength and is lower
than that in the annihilation process approximately by
three orders of magnitude.

The emissivity in the photoneutrino production
process  (\ref{emp1}) in the limit of a strong magnetic field can
be represented as
\begin{equation}
Q_{F}^{(B)} \simeq  \frac{2 C_{+}^{2}}{3 \pi^6} \, \alpha \, I \, G^2_F \, m^2 \, eB \, T^5 \, \ln {(T/m)} ,
\label{Q e gamma}
\end{equation}
where the numerical factor I has the following integral
representation:
$$
 I = \int^{1}_{0} du \int^{u}_{0} d \upsilon  \int^{\infty }_{0} \frac{ dx \, x^4 \, e^{\, -2x} }
 {e^{\, xu} + e^{\, x \upsilon} } \,
\times
$$
\begin{equation}
\times \Big\{
(1 - \upsilon) \left[ (1 - \upsilon)^2 -  (1-u)^2 \right] \,  e^{\, x\,\upsilon } +  \\
 (1+\upsilon )
 \left[ (1 + \upsilon)^2 -  (1 - u)^2 \right]
e^{\, - x \upsilon}
\Big\} \simeq 0{.}09. \nonumber
\end{equation}
Comparison of the derived emissivity with~(\ref{QAB}) shows
a negligible contribution from this reaction to the
neutrino energy losses of a strongly magnetized nondegenerate
plasma.

The neutrino emissivity in the synchrotron emission
process~(\ref{e^mp}) in the strong-field limit was obtained
by Kaminker and Yakovlev (1993):
\begin{equation}
Q^{(B)}_S
=
\frac{1 - 9 / 4 e} {2^{1/4} 9 \, \pi^{9/2}}
C_{+}^{2} \, G^2_F \, (eB)^{17/4} \, T^{1/2} e^{-\sqrt{2eB} / T}  .
\label{QSB}
\end{equation}
As we see from this expression, the emissivity in this
limit is exponentially suppressed by the smallness
of the number density of electrons and positrons at
all Landau levels, except for the ground one. On
these grounds, it is generally concluded that the synchrotron
emission process cannot play a significant
role in the neutrino cooling of a plasma with a strong
magnetic field. However, as will be shown below, this
conclusion is unjustified.

Assuming that the electron--positron pair annihilation
~(\ref{e^+e^-}), is the dominant plasma neutrino cooling
process in the presence of a strong magnetic field, we
can find the characteristic neutrino cooling time $\tau_\nu^{(B)}$
to a temperature $T$. It can be found from the equation
\begin{equation}
\frac{d}{dt}\left[ \frac{eB} {12} T^2 \right] = - Q_A^{(B)},
\label{eq:coolB}
\end{equation}
where the energy density of an electron--positron
plasma whose particles are at the ground Landau
level appears on the left-hand side. Solving this
equation gives
\begin{equation}
\tau^{(B)}_\nu \simeq \frac{8 \, \pi^3}{3 \, \zeta (3) \, C_{+}^{2}} \,
\frac{1}{G_F^2 \, m^2 \, T^3}
\simeq 760~\mbox{s}
\, \left ( \frac{1~\mbox{MeV}}{T} \right )^3  .
\label{eq:tau-nuB}
\end{equation}
Note that the neutrino cooling time depends only on
temperature, because both the plasma energy density
and the energy losses through neutrino emission in
this limit grow proportionally to the magnetic field
strength. It follows from the above estimate that
the characteristic neutrino cooling time for a plasma
with a strong magnetic field must exceed the giant flare
duration $ \tau_{LT} \simeq (200 - 400)$~s by several times. On
these grounds, the authors of the magnetar model
concluded that the plasma neutrino emission could be
neglected (Thompson and Duncan 1995).

Although the neutrino cooling time in this limit
does not depend on magnetic field strength, the neutrino
emissivity~(\ref{QAB}) increases linearly with growing
field, reaching the field-free value~(\ref{QA0}) at $ B \simeq 10^{16} \, G \, (T/m)^4 $.
Thus, the neutrino emissivity in the
annihilation process must be suppressed significantly
in a magnetic field whose strength satisfies the following
inequality:
\begin{equation}
\label{ASup}
4.4 \times 10^{13} \, G \ t^2
\ll B \ll
10^{16} \, G \ t^4  ,
\ \ or \ \
\sqrt{2} \ll x \ll 22 \ t  ,
\end{equation}
where $t = T / m $, $x = \sqrt{2eB}/T$. The lower bound follows from the condition $eB \gg T^2 $.
Note that the
commonly assumed magnetic field strength for magnetars
$ B_M \sim 10^{15}$~G (Duncan and Thompson 1992;
Thompson and Duncan 1993) falls within this range
at a plasma temperature $T \gtrsim m $ typical of a giant
flare. However, the estimate~(\ref{eq:tau-nuB}) for the neutrino
cooling time is valid only if the neutrino emissivity is
defined by the asymptotic expression~(\ref{QAB}). A detailed
analysis shows that this expression is applicable only
in magnetic fields with strengths
\begin{equation}
B \gtrsim 7 \times 10^{15} \, G \ t^2,
\end{equation}
while the emissivity in this process at smaller strengths
can exceed considerably the asymptotic one.

In addition, the neutrino synchrotron emission
process~(\ref{e^mp}) is important for plasma cooling even in a
strong magnetic field. Indeed, the ratio of the emissivity
in this process to the emissivity in the annihilation
reaction is given by the expression
\begin{equation}
\frac{Q_S^{(B)}}{Q_A^{(B)}} =
\frac{\sqrt{2 \pi}} {3 \pi^2 \zeta (3)}
\left( 1 - \frac{9}{4 e} \right) t^2 \, x^{13/2} \, e^{-x}  ,
\label{eq:QS/QAB}
\end{equation}
which has a maximum at $ x_{max} \!=\! 13 / 2$ equal to $Q_S^{(B)} / Q_A^{(B)} \! \simeq 3.5 \ t^2$.  Consequently, at a temperature $T \gtrsim m$
typical of a giant flare, the neutrino synchrotron
emission process contributes significantly
to plasma neutrino cooling in the range of magnetic
field strengths~(\ref{ASup}), where the annihilation process is essentially suppressed.

Thus, our analysis shows that to describe the neutrino
cooling of a plasma emitting an SGR giant flare,
it is insufficient to consider only the electron--positron
pair annihilation~(\ref{e^+e^-}), because the synchrotron neutrino
pair production~(\ref{e^mp}) is a no less important reaction
and both these processes make a comparable
contribution to the plasma energy losses. In addition,
the actual neutrino emissivities in the processes (\ref{e^+e^-})
and (\ref{e^mp}) exceed considerably the asymptotic emissivity~(\ref{QAB}) in the strong-field limit even in magnetic fields
with strengths $B \gtrsim 10^{15}$.

\subsection*{ \bf
LOWER BOUND ON THE MAGNETIC FIELD
STRENGTH OF A MAGNETAR FROM THE NEUTRINO COOLING RATE}

\noindent

\begin{figure}[tb]
\vspace*{-10mm}
\centerline{
\includegraphics[width=200mm]{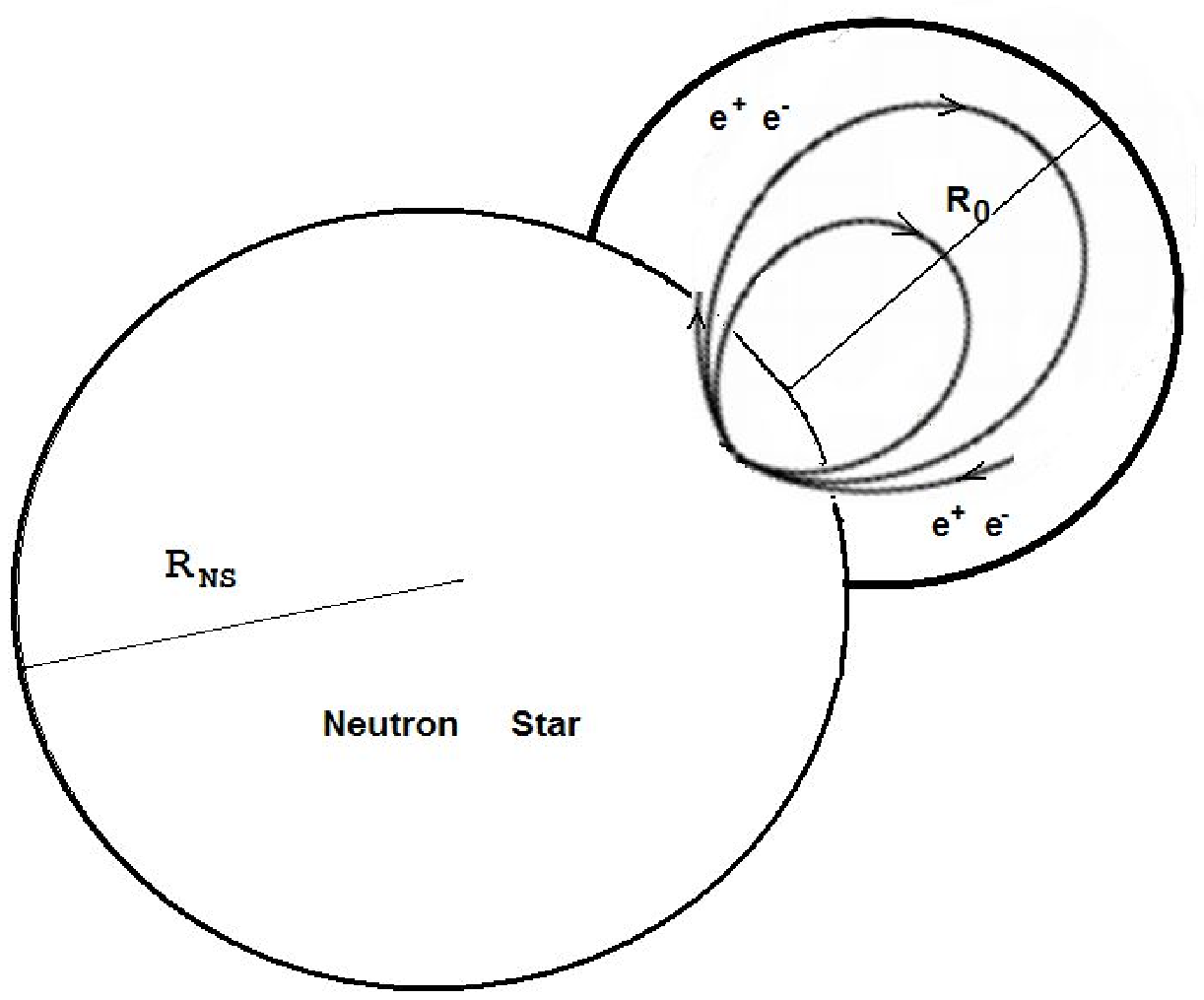}
}
\vspace*{-60mm}
\caption{Scheme of the region of a ball with radius $R_0$
filled with an electron--positron plasma and trapped by
a poloidal magnetic field with field lines closed on the
magnetar crust.}
\label{fig:0}
\end{figure}

In this section, the results obtained above are used
to model the plasma neutrino cooling for an SGR
giant flare. First of all, it should be emphasized that,
although such a plasma is fairly hot, it is essentially
transparent to neutrinos. Consequently, the neutrinos
escape freely from the entire plasma-occupied
volume and the energy lost through neutrino emission
does not depend on the geometry of the emitting
region. To simplify our analysis, we model the fireball
as part of a sphere with radius $R_0$  whose center is
on the magnetar surface~(Fig.~\ref{fig:0}). The distributions
of plasma parameters are assumed to be spherically
symmetric, just as in the paper by Thompson and
Duncan (2001). In our numerical calculations, we
used the following temperature and magnetic field
strength distributions inside the fireball:
\begin{eqnarray}
&& t (z) = t_0 \left ( 1 + z \right )^\gamma,
\label{eq:tz} \\
&& b (z) = b_0 \left ( 1 + z \right )^\beta,
\label{eq:bz}
\end{eqnarray}
where $z = r/R_0$ is the distance from the fireball center
in units of its radius, $ t = T / m$ and $ b = eB/ m^2 $ are the dimensionless temperature and magnetic field strength, respectively. The parameters of the distributions
$t_0, b_0, \beta$ and $\gamma$ allow the neutrino emission to be
described completely. However, in general, we cannot
use the results from Thompson and Duncan (2001),
where these parameters were obtained by comparing
the modeled X-ray emission with the observations of
a giant flare from SGR 1900+14, because the neutrino
emission was completely excluded in this analysis.
Indeed, the plasma at such parameters has only the energy that was observed in this flare as X-ray
emission. In contrast, in the case where the energy
losses of the medium through neutrino emission are
important, the plasma energy must be considerably
higher. Thus, the parameters of the distributions (\ref{eq:tz})
and~(\ref{eq:bz}) must be found by simultaneously modeling
the X-ray emission from the fireball surface and the
neutrino cooling of the medium from its volume and
by comparing the modeling results with the observed
X-ray light curves of SGR giant flares.

Here, we consider a simpler model. The electron--positron
plasma is assumed to have an energy $E_{tot}$
that is expended in cooling by neutrino emission and
an SGR X-ray flare with an energy $E_{LT}$. In this case,
the energy balance equations can be written as
\begin{eqnarray}
\label{eta ELT}
E_{tot} = \eta \, E_{LT}
=
2 \pi R_0^3 \int\limits_0^1
\Big( U_{e^{\pm}}(z) + U_\gamma(z) \Big) \, z^2 \, dz  ,
\\
\label{eta-1 ELT}
E_\nu =
\left ( \eta - 1 \right ) E_{LT}
=
2 \pi R_0^3 \, \tau_{LT}  \int\limits_0^1 Q_{\nu}(z) \, z^2 \,dz  ,
\end{eqnarray}
where $\eta = E_{tot} / E_{LT} $ the plasma is assumed to
be composed of electrons, positrons, and photons, $U_{e^{\pm}}(z)$ and $U_\gamma(z)$
are the local energy densities of
these particles, $Q_{\nu}(z)$ is the neutrino emissivity, and
the right-hand side of the second equation gives the
total energy loss $E_\nu$ through neutrino cooling in the
giant-flare time $\tau_{LT}$.

\subsection*{\it
Analytical Neutrino Emission Model
in the Asymptotic Limit of a Strong Magnetic Field}

\noindent

\begin{figure}[tb]
\centerline{
\includegraphics[width=70mm]{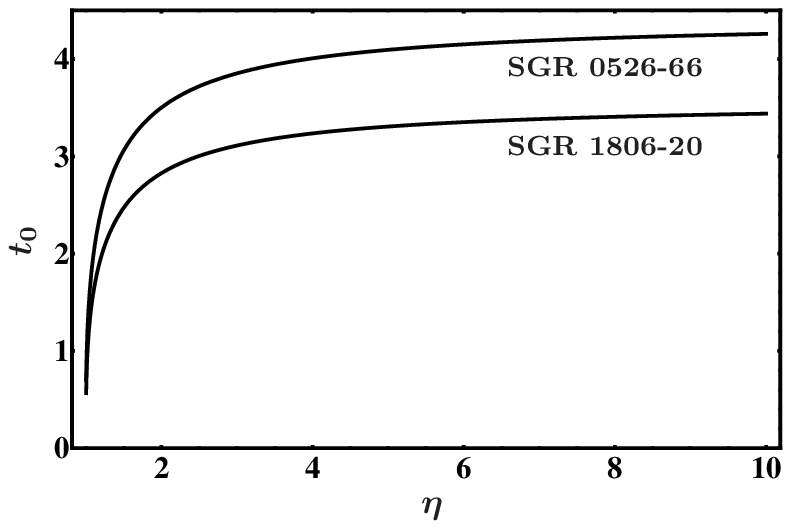}
\hfil %
\includegraphics[width=70mm]{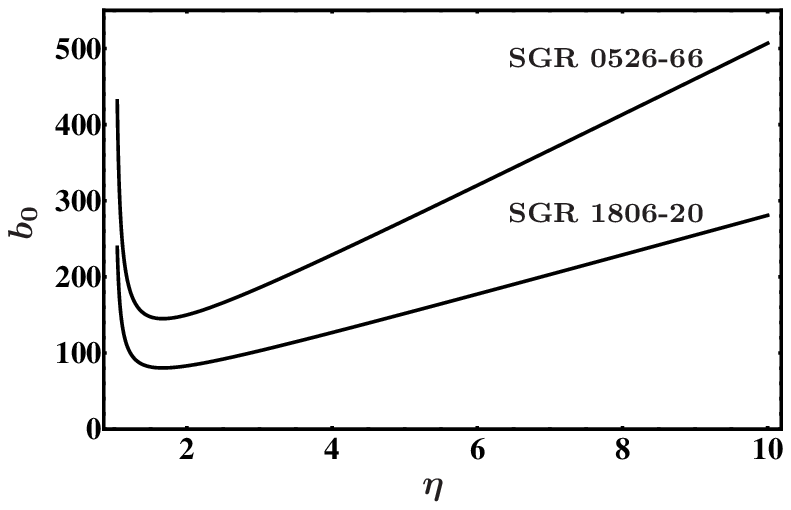}}
\caption{Dependences of the temperature $t_0 (a)$ and magnetic field strength $b_0 (b)$ at the fireball center on parameter ç
corresponding to the analytical solution. The lines are drawn for $R_0 = 10$ km and the parameters $ \gamma = -1/2$  and  $ \beta = - 3$. }
\label{fig:1}
\end{figure}

As was shown above, the asymptotic expressions
for the neutrino emissivities derived in the limit of a
strong magnetic field cannot be used to describe the
neutrino emission of an SGR giant flare. However,
we will consider this limiting case, because it admits
an analytical solution and provides an insight into
the main features of neutrino cooling in the magnetar
model. In this case, the neutrino energy losses are
determined only by the electron--positron pair annihilation~(\ref{e^+e^-}):
\begin{equation}
Q_{\nu} = Q^{(B)}_A  ,
\end{equation}
whose emissivity is given by Eq. (11). The plasma
energy density in this limit is determined by electrons
and positrons at the ground Landau level:
\begin{equation}
U_{e^{\pm}} + U_\gamma
\simeq U_{e^{\pm}} \simeq \frac{\, m^4}{12} \, b \, t^2  .
\end{equation}
The energy balance equations (\ref{eta ELT})  and (\ref{eta-1 ELT}) in this
case can be represented as
\begin{eqnarray}
\label{eta ELT B}
&& \hspace{2cm} \eta \, E_{LT}
=
\frac{\pi}{6} \, J \!\left(\beta + 2 \gamma \right) \, m^4 R_0^3 \, b_0 \, t_0^2  ,
\\
\label{eta-1 ELT B}
&& \left ( \eta - 1 \right ) E_{LT}
=
\frac{\zeta (3) \, C_{+}^{2}} {24 \pi^2} \, J \!\left(\beta + 5 \gamma \right)
\, G^2_F \, m^9 \, \tau_{LT} \, R_0^3 \, b_0 \, t_0^5  ,
\end{eqnarray}
where the function $J(\delta)$ is defined by the integral
\begin{equation}
J(\delta)
=
\int\limits_0^1 (1 + z)^\delta z^2 dz
=
\frac{2^{\delta + 1} (\delta^2 + \delta +2) - 2} {(\delta + 1) (\delta + 2) (\delta + 3)}  ,
\nonumber
\end{equation}
and the previously introduced temperature~(\ref{eq:tz})  and
magnetic field strength (\ref{eq:bz}) distributions are used.
The solution of this system of equations can be represented
as
\begin{eqnarray}
\label{t0}
&& t_0(\eta)
=
\left( \frac{4 \pi^3} {\zeta (3) \, C_{+}^{2}}
\, \frac{J \!\left(\beta + 2 \gamma \right)} {J \!\left(\beta + 5 \gamma \right)}
\, \frac{1} {G^2_F \, m^5 \, \tau_{LT}}  \,  \frac{\eta - 1} {\eta}
\right)^{1/3}  ,
\\
\label{b0}
&& b_0(\eta)
=
\frac{3 \left( \zeta (3) \, C_{+}^{2} \right)^{1/3}} {2^{1/3} \, \pi^3}
\frac{J^{2/3} \!\left(\beta + 5 \gamma \right)} {J^{5/3} \!\left(\beta + 2 \gamma \right)}
\, \frac{G_F^{4/3} \, E_{LT} \, \tau_{LT}^{2/3}} {m^{2/3} R_0^3}
\frac{\eta^{5/3}} {(\eta - 1)^{2/3}}  .
\end{eqnarray}
Analysis of this solution shows that $t_0(\eta)$ and $\beta$
depend very weakly on the exponents $b_0(\eta)$ and $\gamma$,
respectively.
In addition, in the magnetar model (Thompson
and Duncan 2001), the change in temperature
inside the fireball is assumed to be fairly small —
from the isothermal case ($ \gamma = 0 $)
to the case of $\gamma = -1$. Of particular interest is $ \gamma = -1/2 $
corresponding to an arbitrary exponent $\beta$.
Since a sharp decrease
in magnetic field strength in the fireball would
lead to excessively intense cooling of its outer layers
through neutrino emission, we will assume that the
field decreases no faster than the dipolar law $\beta = -3$.
 Under these assumptions, the following approximate
expressions can be used to estimate the temperature
$T_0 \simeq   0.51~ MeV \, \cdot t_0$ and magnetic
field strength $B_0 \simeq 4.4 \times 10^{13}~ G  \, \cdot b_0 $ at the fireball center:
\begin{eqnarray}
\label{t0sim}
&& \hspace{1.2cm} t_0(\eta) \simeq
4.4 \, (1 - 0.6 \gamma) \, \frac{1}{\tau_{100}^{1/3}} \left( \frac{\eta - 1} {\eta} \right)^{1/3}  ,
\\
\label{b0sim}
&& b_0(\eta) \simeq
2.1 \, (1 - 0.5 \beta + 0.3 \beta^2)
\, \frac{E_{44} \tau_{100}^{2/3}} {R_{10}^3}  \,  \frac{\eta^{5/3}} {(\eta - 1)^{2/3}}  ,
\end{eqnarray}
where $ \tau_{100} = \tau_{LT} / 100$~s, $E_{44} = E_{LT} / 10^{44} $~erg and $_{10} = R_0 / 10$~km. As we see from these expressions,
the solution $t_0 (\eta)$ grows rapidly with plasma energy $E_{tot} = \eta \, E_{LT}$ in the case of relatively small neutrino energy losses, when $\eta \simeq 1$.
Thus, small neutrino
energy losses are possible only in a fairly cold plasma.
In the region where neutrino cooling dominates, when $\eta \gg 1 $,
the solution $t_0(\eta)$ reaches a constant. The
existence of this limiting temperature stems from the
fact that the change in fireball temperature through
the emission of neutrinos and photons in the flare
time $\tau_{LT}$ is neglected in the model under consideration.
The solution $b_0(\eta)$ contains a divergence at $\eta = 1$ which physically corresponds to the impossibility of completely removing the energy losses through
neutrino emission. It passes through a minimum at $\eta = 5 / 3$
and reaches the asymptotics $\eta$ as $b_0(\eta) \sim \eta$
increases further in the range of energies where the
neutrino energy losses dominate. Figure ~\ref{fig:1} presents
the corresponding solutions at $\gamma = -1/2$ and $\beta = -3$
for the following giant flares:
\begin{eqnarray}
\label{SGR1}
&& SGR~0526-66: \ \ E_{44} = 3.6, \ \tau_{100} = 2.0,
\\
\label{SGR2}
&& SGR~1806-20: \ \ E_{44} = 1.3, \ \tau_{100} = 3.8.
\end{eqnarray}
We do not discuss the flare from SGR 1900+14,
because it has characteristics at the LT state similar
to those of SGR 1806--20.

Let us separately consider another important peculiarity
of the solution obtained. If we fix the total
plasma energy $E_{tot} = \eta E_{LT}$ and assume that the
temperature $t_0 > t_0(\eta)$ then, in view of Eq.~(\ref{eta ELT B}), the
magnetic field strength $b_0$ must be lower than the corresponding
solution $b_0(\eta)$. It is easy to see that, in this
case, the energy losses by such a medium  $E_\nu$
through neutrino emission that are defined by the right-hand
side of Eq.~(\ref{eta-1 ELT B}) will increase, i.e., they will become
greater than $(E_{tot} - E_{LT})$.
In contrast, for  $t_0 < t_0(\eta)$
the value of $b_0$ must be higher than $b_0(\eta)$
and the
neutrino energy losses will decrease. Thus, the solution
~(\ref{t0}) defines the maximum temperature, while
the solution ~(\ref{b0}) defines the minimum magnetic field
strength at which the neutrino energy losses leave an
energy in the plasma no less than $E_{LT}$
observed in the
photon emission. In addition, since $t_0(\eta)$
has a global maximum and $b_0(\eta)$ has a global minimum, the upper
and lower bounds, respectively, on the admissible
temperature and the magnetic field strength at the
fireball center correspond to them:
\begin{eqnarray}
\label{T0max}
&& T_0^{(max)} \simeq
2.2 \, MeV
\, (1 - 0.6 \gamma) \, \frac{1}{\tau_{100}^{1/3}}  ,
\\
\label{B0max}
&& B_0^{(min)}
\simeq
2.8 \times 10^{14} \, G
\left( 1 - 0.5 \, \beta + 0.3 \, \beta^2 \right) \frac{E_{44} \, \tau_{100}^{2/3}} {R_{10}^3}  ,
\end{eqnarray}
They are needed for the energy emitted by the plasma
in photons to be no less than $E_{LT}$.
As was noted
above, the existence of a maximum temperature of the
medium stems from the fact that its change through
the emission of photons and neutrinos in the flare time
$\tau_{LT}$ is disregarded in the model under consideration.
It follows from the condition (\ref{B0max}) that, depending on
the magnetic field strength, a plasma with an arbitrary
energy $E_{tot}$ can emit an energy in photons no greater
than
\begin{equation}
E_{LT}^{(max)}
\simeq
10^{44} \, erg \, \frac{B_{15} \, R_{10}^3} {\tau_{100}^{2/3}}  ,
\end{equation}
where $B_{15} = B_0 / 10^{15} G $ and $\beta = -3$. As we see from
this expression, the fireball radius $R_0$
is the most
important parameter defining the admissible energy
release in photons, while the reasonable energy release
itself is close to the typical energy of SGR giant
flares.

It should be noted that a modulation of the
emission intensity by the neutron star spin period
(Mazets et al. 1979; Ibrahim et al. 2001; Mereghetti
et al. 2005; Frederiks et al. 2007) is clearly traceable
for all of the known SGR giant flares at the LT
stage. The presence of such pulsations leads to the
conclusion that the size $R_0$ of the plasma-occupied
region must be close to the neutron star radius $R_{NS}$.
Under this assumption and for a standard value of $R_{NS} \simeq 10$
km and a typical energy $E_{LT} \sim 10^{44}$ erg
of an SGR giant flare at the LT stage, the magnetic
field strength must be $B_0 \gtrsim 10^{15}$~G.
It is interesting
to compare this value with the upper bound on the
magnetic field strength that follows from an estimate
of the magnetic dipole losses for a magnetar:
\begin{equation}
\label{BMD}
B_{MD}
\simeq
2.1 \times 10^{15} \, G
\ \frac{1} {\sin\theta} \, \frac{M_{1.4}^{1/2}} {\widetilde R_{10}^2} \left( P_{10} \, \dot P_{-10} \right)^{1/2}  .
\end{equation}
Here, $M_{1.4} = M_{NS} / 1.4 M_\odot$, $\widetilde R_{10} = R_{NS} / 10 km $, $P_{10} = P / 10 s $,
$\dot P_{-10} = \dot P / 10^{-10}$, where $M_{NS}$, $P$,  and  $ \dot P $
are the neutron star mass, period, and spindown
rate, respectively, $\theta$ is the angle between the angular
velocity and magnetic moment vectors. It is easy to
see that these bounds almost coincide at $R_0 \simeq R_{NS}$.
Thus, the above estimates leave open only a narrow
range of magnetar magnetic field strengths even
in the simplified model that grossly underestimates
the energy losses of the medium through neutrino
emission and, consequently, the minimum magnetic
field strength $B_0^{(min)}$.

\subsection*{\it Modeling the Neutrino Cooling of SGR 0526--66
and SGR 1806--20}

\noindent

In the previous section, we studied an analytical
model for the neutrino cooling of a plasma producing
an SGR giant flare that is based on the asymptotic
expressions for the plasma energy density and neutrino
emissivity derived in the limit of very strong
magnetic fields. In this section, we will use the expressions
valid for an arbitrary magnetic field strength
for a more realistic modeling of the neutrino cooling
process.

Since the photon mean free path in the plasma
under consideration is small, the electrons, positrons,
and photons are in local thermodynamic equilibrium
and, hence, the equilibrium distribution functions can
be used to describe them. In this case, the plasma
energy density will be defined by the expressions
\begin{eqnarray}
&& \hspace{4cm} U_\gamma = \frac{\pi^2 m^4} {15} \, t^4  ,
\\
&& U_{e^{\pm}}
=
\frac{m^4}{\pi^2} \, b \left[
\int\limits_0^\infty \frac{(x^2 + 1)^{1/2}} {e^{\sqrt{x^2 + 1} / t} + 1} \, dx
+ 2 \sum\limits_{n = 1}^\infty \int\limits_0^\infty \frac{(x^2 + 2bn + 1)^{1/2}} {e^{\sqrt{x^2 + 2bn + 1} / t} + 1} \, dx
\right]  .
\end{eqnarray}
As was shown above, the electron--positron pair annihilation
~(\ref{e^+e^-}) and the neutrino synchrotron emission
~(\ref{e^mp}) are the main neutrino cooling processes for
the plasma under consideration:
\begin{equation}
Q_{\nu} = Q_A + Q_S.
\end{equation}
Here, for the neutrino emissivity of the annihilation
process $Q_A$, we used the interpolation formula
(Kaminker et al. 1992)
\begin{eqnarray}
&& \hspace{-2.6cm} Q_A
=
\frac{G_F^2 m^9} {\pi^4} \Bigg\{
\left[ \left( C_{+}^{2} + C_{-}^{2} \right) \left( \frac{1}{2} + \frac{15}{8} \, t \right) t^3
+  C_{+}^{2} \, P(t) \right] F(t,b)
\\
\nonumber
&& + \left[ C_{+}^{2} + C_{-}^{2} + \left( C_{+}^{2} - C_{-}^{2} \right) \frac{b} {1 + b} \right]
\frac{b^2 \, S(t)} {12 \, (1 + b)}
\Bigg\} \exp\left(\! - \frac{2}{t} \,\right)  ,
\\
\nonumber
&& P(t)
=
t^4 \left( 1 + 3.581\, t + 39.64 \, t^2 + 24.43 \, t^3 + 36.49 \, t^4 + 18.75 \, t^5 \right)  ,
\\
\nonumber
&& S(t)
=
t \left( 1 + 1.058\, t + 0.6701 \, t^2 + 0.9143 \, t^3 + 0.472 \, t^4 \right)  ,
\\
\nonumber
&& F(t, b) = \frac{1} {R_1 R_2 R_3}  ,
\ \ R_i = 1 + c_i \, \frac{b}{t^2} \, \exp\left( \frac{\sqrt{2 b}} {3 t} \right)  ,
\end{eqnarray}
where $C_{+}^{2} \simeq 1.675$, $ C_{-}^{2} \simeq 0.175 $ the constants $ c_1 \simeq 3.106 \times 10^{-6} $, $ c_2 \simeq 1.491 \times 10^{-3} $, $ c_3 \simeq 4.839 \times 10^{-6} $.  For the emissivity of the synchrotron neutrino pair
production reaction, we used the interpolation expression
from Kaminker and Yakovlev (1993)
\begin{eqnarray}
&& \hspace{4cm} Q_S
=
\frac{G_F^2 m^9} {120 \, \pi^3} \, N_{e^{\pm}}
\left( C_{+}^{2} F_{+} - C_{-}^{2} F_{-} \right)  ,
\\
\nonumber
&& \hspace{4cm} N_{e^{\pm}}
=
\frac{2 \, m^3}{\pi^2} \, b
\sum\limits_{n = 1}^\infty \int\limits_0^\infty \frac{dx} {e^{\sqrt{x^2 + 2bn + 1} / t} + 1}  ,
\\
\nonumber
&& \hspace{-0.7cm} F_{+}
=
\frac{b^6} {(1 + a_1 b)^3}  \left( 1 + \frac{a_2 \, t} {(1 + a_1 b)^{1/2}} \right)^{6}
\!\left( 1 + \frac{ a_3 \, t \, b} {(1 + a_1 b)^{3/2}} \right)^{-5}
\!\left[ 1 + \ln \left( 1 + \frac{a_4 \, t \, b} {(1 + a_5 b)^{3/2}} \right) \right]  ,
\\
\nonumber
&& \hspace{3.8cm} F_{-}
=
\frac{b^6 (1 + a_6 \, t)^6} {(1 + a_1 b)^6}
\left( 1 + \frac{ a_3 \, t \, b} {(1 + a_1 b)^{3/2}} \right)^{-5}  ,
\end{eqnarray}
where $N_{e^{\pm}}$ is the number density of electrons and
positrons at all Landau levels, except for the ground
one. Here, the constants $ a_1 \simeq 0.955 $, $ a_2 \simeq 9.439 $, $ a_3 \simeq 23.31 $, $ a_4 \simeq 0.26 $,
$ a_5 \simeq 0.168 $, $ a_6 \simeq 0.971$.

\begin{figure}[tb]
\centerline{
\includegraphics[width=70mm]{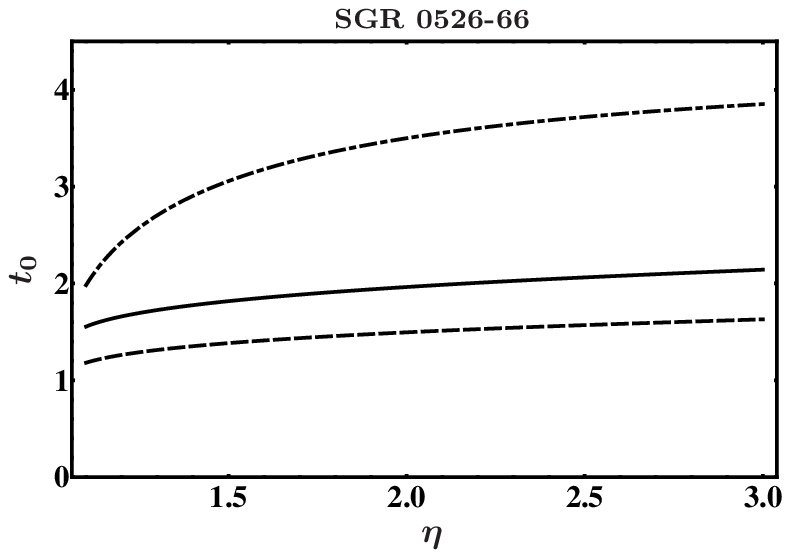}
\hfil %
\includegraphics[width=70mm]{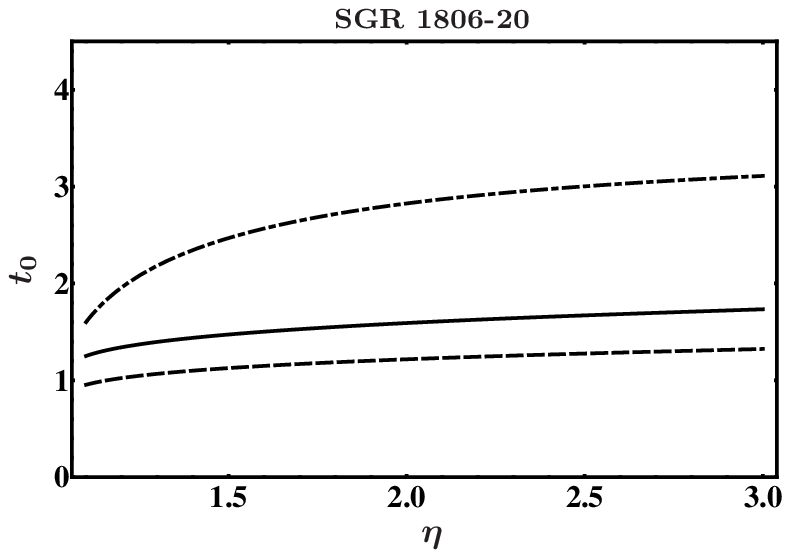}
}
\caption{Temperature $ t_0$  at the fireball center versus parameter ç for the SGR 0526--66 (a) and SGR 1806--20 (b) flares at
$R_0 = 10$ km and $ \beta = - 3$. The solid, dashed, and dash–dotted lines correspond to
$ \gamma = -1/2$, $ \gamma = 0$, and the analytical solution
at $ \gamma = -1/2$, respectively.}
\label{fig:2}
\end{figure}

\begin{figure}[tb]
\centerline{
\includegraphics[width=70mm]{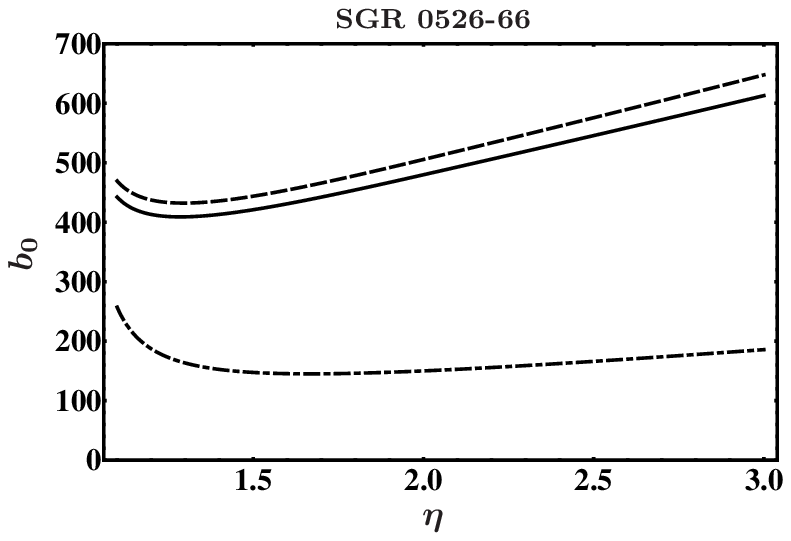}
\hfil %
\includegraphics[width=70mm]{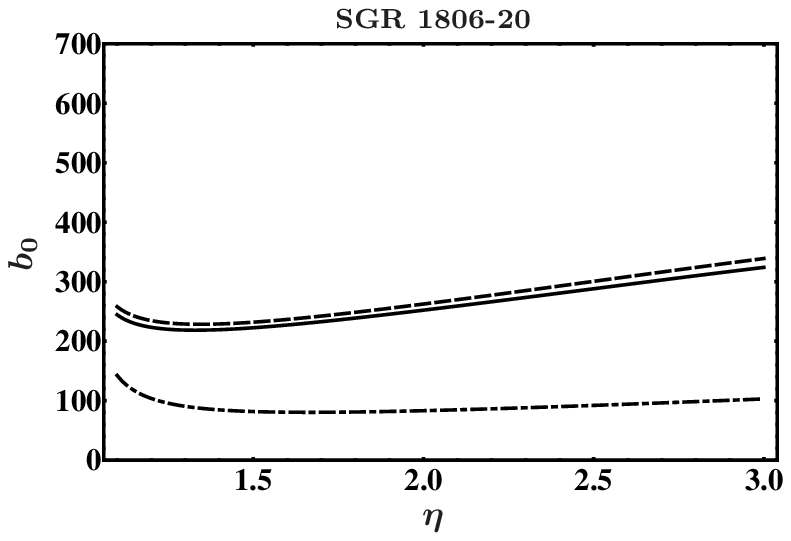}
}
\caption{Magnetic field strength b0 at the fireball center versus parameter ç for the SGR 0526--66 (a) and SGR 1806--20 (b) flares at $R_0 = 10$ km and  $ \beta = - 3$. The solid, dashed, and dash–dotted lines correspond to $ \gamma = -1/2$, $ \gamma = 0$, and the analytical solution at $ \gamma = -1/2$,  respectively.}
\label{fig:3}
\end{figure}

We modeled neutrino cooling for giant flares from
SGR 0526--66 and SGR 1806--20 with the characteristics
(\ref{SGR1}) and (\ref{SGR2}). As was noted above, the
SGR 1900+14 flare at the LT stage is similar in
its properties to SGR 1806--20 and below we do
not separate them. Our numerical calculations confirm
all of the qualitative conclusions reached in the
previous section for the analytical model of neutrino
cooling, but the numerical values of the plasma temperature
and magnetic field strength change significantly.
Below, we present the results for the case
of $\beta = -3 $ which corresponds to a dipole magnetic
field configuration in the fireball. The law of change
in temperature was chosen with an exponent characteristic
of the magnetar model, $\gamma = - 1/2$ which
was compared with the isothermal case of $\gamma = 0$.
The fireball radius was chosen to be  $R_0 = 10$~km, which
corresponds to the standard radius of a neutron star.
The results of our numerical solution of the system of
equations (\ref{eta ELT}, (\ref{eta-1 ELT})  for the SGR giant flares under
consideration are presented in Figs. ~\ref{fig:2} and \ref{fig:3}. As we
see from the plots, the general trend in the behavior
of the temperature and magnetic field strength as a
function of the total plasma energy $E_{tot} = \eta E_{LT}$
is the
same as that in the analytical model. However, the
plasma being analyzed turns out to be colder and the
magnetic field strength in it must be higher by several
times. Note that the dependence of the solutions
on parameters $\gamma$ and $\beta$
also remains similar to the
analytical model (see Eqs.(\ref{t0sim}) and (\ref{b0sim})). Thus, the
calculated magnetic field strength in the fireball is
virtually independent of the law of change in temperature,
as is demonstrated in  Fig.~\ref{fig:3}. We emphasize
once again that the derived field strength at fixed total
plasma energy is minimally possible for the emission
of the observed giant-flare energy $E_{LT}$.
Note also that
the global minimum of the magnetic field strength for
the SGR giant flares in question takes place at $\eta \simeq 1.3$
instead of $\eta = 5/3 $
in the analytical model and,
hence, the energy losses through neutrino emission
are reduced approximately by half. However, the
losses remain fairly large and their reduction requires
a great increase in magnetic field strength even in this
case.

\begin{figure}[tb]
\centerline{
\includegraphics[width=70mm]{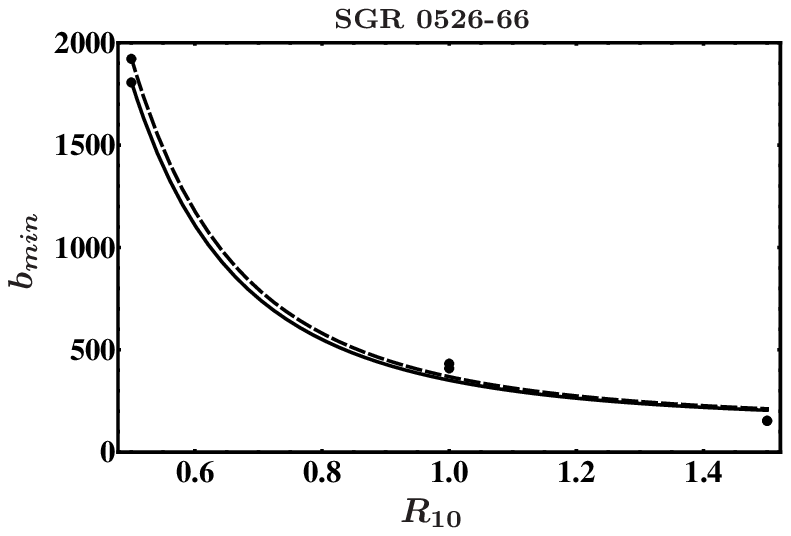}
\hfil %
\includegraphics[width=70mm]{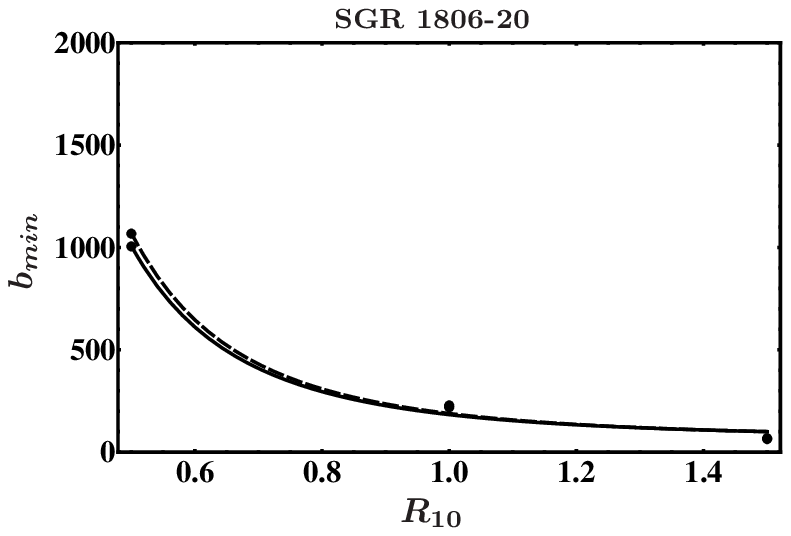}
}
\caption{Minimum magnetic field strength $b^{min}_0$ versus radius $R_{10}$ of the plasma-occupied region for the SGR 0526--66 (a) and
SGR 1806--20 (b) flares. The filled circles correspond to the calculated values of  $b^{min}_0$ at $R_0 = 5, 10, 15$ km and $ \gamma = -1/2$,
 $ \beta = 0 $.The solid and dashed lines represent the fits to this dependence at $ \gamma = -1/2$  and $ \gamma = 0$, respectively.}
\label{fig:4}
\end{figure}

We separately investigated the dependence of the
solutions obtained on the plasma-occupied fireball
radius $R_0$. In contrast to the analytical model where
the plasma temperature did not depend on this parameter
but was determined only by the giant-flare
duration, our numerical calculation showed that the
temperature of the plasma-occupied region decreases
with its increasing sizes. The dependence of the
magnetic field strength on $R_0$ remains similar to the
analytical model, but the law $b_0 \sim R_{10}^{-3}$  is replaced
by a slightly faster decrease in strength. Figure ~\ref{fig:4}
presents the numerically calculated global minimum
of themagnetic field strength $b_{min} \equiv b_0^{(min)}(\eta)$
at $ R_0 =$ 5~, 10  and $15$~km.
It is well fitted by the formulas
\begin{equation}
\nonumber
SGR~0526-66: \ \ b_{min} \simeq 220 \, R_{10}^{-3} + 140  ;
\ \ \ SGR~1806-20: \ \ b_{min} \simeq 120 \, R_{10}^{-3} + 60   ,
\end{equation}
which are valid both for the isothermal case
$(\gamma = 0)$ and for the case of $\gamma = - 1/2$. As we see from
the plots, the minimum magnetic field strength near
$R_0 \simeq 5 $ km exceeds considerably $B_0^{(min)}  \sim 10^{16}$~G.
Therefore, the situation where the plasma occupies
a fairly extended region $R_0 \gtrsim 10 $ km is of greatest
interest in our analysis. In this case, the minimum
magnetic field strength at the fireball center and the
upper bound on the field strength obtained from our
estimate of the magnetar magnetic dipole losses can
be represented as
\begin{eqnarray}
\label{B1}
&& \hspace{-1.5cm}
SGR~0526-66: \hspace{0.7cm}  B_0^{(min)} \simeq 2 \, R_{10}^{-3} \times 10^{16} \, G  ,
\hspace{0.5cm} B_{MD} \simeq 2 \, \widetilde R_{10}^{-2} \times 10^{15} \, G  ;
\\
\label{B2}
&& \hspace{-1.5cm}
 SGR~1806-20: \hspace{0.7cm}  B_0^{(min)} \simeq  R_{10}^{-3} \times 10^{16} \, G   ,
\hspace{0.8cm}  B_{MD} \simeq (2 - 6) \, \widetilde R_{10}^{-2} \times 10^{15} \, G  ;
\\
\label{B3}
&& \hspace{-1.5cm}
SGR~1900+14: \hspace{0.5cm}  B_0^{(min)} \simeq  R_{10}^{-3} \times 10^{16} \, G ,
\hspace{0.8cm}  B_{MD} \simeq (2 - 3) \, \widetilde R_{10}^{-2} \times 10^{15} \, G  ,
\end{eqnarray}
where $B_0^{(min)}$ for SGR 1900+14 is the same as
that for the SGR 1806--20 giant flare. To estimate
the strength $B_{MD}$, we used the parameters
$M_{NS} = 1.4 M_\odot$ and $\theta = \pi / 4$
and the following periods
and spindown rates:
SGR 0526--66:
$P_{10} \simeq 0.81 $, $ \dot P_{-10} \simeq 0.65$;
SGR 1806--20:
$P_{10} \simeq 0.756 $, $ \dot P_{-10} \simeq 0.8 \div 8$;
SGR 1900+14:
$P_{10} \simeq 0.515 $, $ \dot P_{-10} \simeq 0.5 \div 1.4$,
(Mereghetti 2008).
As was discussed above, the
modulation of the X-ray intensity for known flares
by the magnetar spin period leads to the conclusion
that the fireball radius $R_0$ cannot differ significantly
from the neutron star radius $R_{NS}$. As follows from the
estimates (\ref{B1}) -- (\ref{B3}), under the condition $R_0 \simeq R_{NS}$
the upper bound on the magnetic field strength of
the magnetars under consideration turns out to be
lower than the minimally possible value required for
the suppression of neutrino emission.

\subsection*{CONCLUSIONS}

In this paper, we estimated the energy losses
of a nondegenerate relativistic $(T \gtrsim m)$
electron--positron plasma through neutrino emission based on
the magnetar model of an SGR giant flare (Thompson
and Duncan 1995, 2001). In the absence of a
magnetic field, the plasma energy losses through
neutrino emission were shown to be too large to
provide the observed energy release at the LT stage
of an SGR giant flare. It follows from our analysis of
the neutrino processes considered that in the case of
a strongly magnetized plasma $e B \gg m^2$ important
for the magnetar model, not only the electron--positron annihilation into a neutrino pair ~(\ref{e^+e^-}) but also
the neutrino synchrotron emission process~(\ref{e^mp}), which
is usually neglected, make a major contribution to
the neutrino energy losses. The plasma neutrino
emissivities in these processes were shown to be
significant even in the case of fairly strong magnetic
fields $B \gtrsim 10^{15} $~G. Thus, when the X-ray emission
of an SGR giant flare is modeled, the plasma energy
losses through neutrino emission should be properly
taken into account, which was not done by Thompson
and Duncan (1995, 2001).

To investigate the main features of the fireball
neutrino cooling, we considered a simple analytical
model. It clearly shows that the photon emission
at an arbitrary plasma energy cannot exceed some
maximum value dependent on the size of the plasma occupied
region and the magnetic field strength in the
plasma. Thus, the minimally possible magnetic field
strength of a magnetar that provides sufficient suppression
of its neutrino emission at the long tail stage
can be found for the observed energy of a giant flare.
We numerically modeled the neutrino cooling of the
SGR 0526--66, SGR 1806--20, and SGR 1900+14
giant flares, including all of the neutrino reactions
important for this process. The lower bound on the
magnetic field of these objects corresponding to the
energy observed in photons at the LT stage of SGR
giant flares was shown to disagree with the upper
bound from our estimate of their magnetic dipole
losses. Consequently, the magnetar model of an
SGR giant flare considered here cannot provide the
energetics observed at the long tail stage in a fairly
wide range of parameters.

We modeled neutrino cooling under the simplifying
assumptions that the fireball temperature and
sizes did not change in the time of an SGR giant flare.
Allowance for the evolution of these characteristics
must lead to a reduction in the energy losses of the
medium through neutrino emission compared to the
model considered. However, it is hard to expect
that the minimally possible magnetic field strength
required for the suppression of the fireball neutrino
emission can be significantly lower than the estimates
obtained here and can become equal to the upper
bounds following from our estimate of the magnetic
dipole losses for magnetars. Note also that the magnetar
model is not without other contradictions either.
In particular, an important problem (Malov and
Machabeli 2006) that arises in attempting to explain
the existence of radio emission detected from SGRs
and AXPs (Malofeev et al. 2005) should be pointed
out.

In conclusion, note that the magnetar model
considered here was further developed by Lyutikov
(2006), Beloborodov and Thompson (2007), and
Beloborodov (2009). These authors used the more
realistic magnetohydrodynamic approach to describe
the magnetar corona and revealed the effect of additional
plasma heating due to the magnetic field
energy. Using this approach to describe SGR giant
flares at the LT stage could partially solve the problem
of energy deficiency discussed here.

\subsection*{ACKNOWLEDGMENTS}

We wish to thank G.S. Bisnovatyi-Kogan,
S.I. Blinnikov, and N.V.Mikheev for constant interest
in the work and fruitful discussions of the results. We
are grateful to A.D.Kaminker, I.F.Malov, S.B. Popov,
D.A. Rumyantsev, M.V. Chistyakov, A.I. Tsygan, and
D.G. Yakovlev for helpful discussions and valuable
remarks. We are also grateful to the referee for the
remarks whose allowance improved the paper. The
study was supported by the "Scientific and Scientific--Pedagogical Personnel of Innovational Russia" Federal
Goal-Oriented Program for 2009--2013 (State
contract no.  P2323) and, in part, by the "Development
of the Scientific Potential of Higher School"
Program of the Ministry of Education and Science of
the Russian Federation (project no.  2.1.1/510).


\end{document}